# Payload-Size and Deadline-Aware Scheduling for Time-critical Cyber Physical Systems

**Marcus Haferkamp, Benjamin Sliwa, Christoph Ide and Christian Wietfeld**
Communication Networks Institute
TU Dortmund University
44227 Dortmund, Germany
e-mail: {Marcus.Haferkamp, Benjamin.Sliwa, Christoph.Ide, Christian.Wietfeld}@tu-dortmund.de

*Abstract*—High transfer speeds, low latencies and a widespread availability qualify Long Term Evolution (LTE) for various applications and services in the field of Human-to-Human (H2H) as well as fast growing Vehicle-To-X (V2X) and Cyber Physical Systems (CPS) communications. As a result, a steady growth of mobile data traffic causing an increasing interaction of different traffic classes can be observed. In order to ensure timely transmissions of time-critical data in the future, we propose the novel Payload-Size and Deadline-Aware (PayDA) scheduling approach and compare its performance regarding the compliance with deadlines with those of other common packet scheduling mechanisms. The performance analysis is done with the complex and open-source LTE simulation environment LTE-Sim. The results show that the average latency can be reduced by the factor of 20 and the mean goodput can be enhanced by a factor of about 3.5 for a high miscellaneous data traffic. In case of a heavy homogeneous and time-critical data traffic the mean Deadline-Miss-Ratio (DMR) can be decreased by about 35%.

## I. INTRODUCTION

In the course of the introduction of LTE and future 5th Generation Mobile Networks (5G) a wider range of new application types with various requirements has been and will be established (e.g., in vehicular or smart grid environments) leading to a constant growth of demands and utilization of the existing mobile network infrastructure. As a result, there is a negative impact on the compliance with Quality of Service (QoS) requirements during the transmission of data. The most important QoS-criterion for (safety-related) real-time applications is the adherence to deadlines. While some existing real-time strategies propose to meet given deadlines, there are still no efficient techniques for mobile networks established which take deadlines as well as payload sizes into account. Resulting from this, there are many open research questions left which have to be solved to reach the targets of future 5G mobile networks. Hence, the focus of this paper is to propose a novel payload-size and deadline-aware scheduling mechanism for Resource Block (RB) allocation regarding downlink transmissions, inspired by the well-known Earliest Deadline First (EDF) scheduler, to close this gap. Fig. 1 exemplifies the prioritization of User Equipments (UEs) in a simple mobile cell scenario due to deadlines regarding the data transmission and the buffer size of the Medium Access Control (MAC) queue. The structure of this paper is as follows: First of all, the related work is discussed in Sec. II which is further followed by the presentation of the proposed PayDA scheduling strategy in Sec. III. In Sec. IV, we introduce the simulation environment LTE-Sim [1] used to analyze the impact of EDF and PayDA on various key performance indicators of data transmission (e.g., DMR and latency) for different data traffic patterns in LTE. Finally, detailed results of a simulation study are presented in Sec. V which clarify that the proposed PayDA scheduling scheme drastically reduces the average latency of data transmissions, while at the same time being at least as efficient as the EDF algorithm according to the DMR. As an overall conclusion, we outline that the results of this paper emphasize the need for a multi-criteria real-time scheduling strategy in mobile networks.

## II. RELATED WORK

Besides the challenges of wired communication systems, mobile communication systems have to accomplish additional effects regarding the users' mobility and thus permanently changing channel conditions. For this reason, there are many different resource scheduling mechanisms in LTE with manifold objectives (e.g., maximization of data rates, high degrees of fairness regarding the allocation of RBs), which take the channel conditions in terms of the spectral efficiency into consideration. In [2], the authors give a detailed overview of several types of common scheduling strategies and provide an assessment about the suitability of these procedures for use with different kinds of H2H data traffic in LTE downlink transmissions. A novel, two-stage scheduling procedure with respect to QoS support in the downlink of LTE cellular networks is analyzed for high-rate and time-critical multimedia data traffic in [3]. The simulation results reveal quite low DMRs, but up to the factor of 3 lower data rates for best effort services than other common real-time schedulers. A comprehensive analysis of the effects due to interdependences of time-critical H2H and non-real-time Machine-to-Machine (M2M) data traffic on QoS is given in [4]. With respect to M2M communication, in [5] a scheduler is proposed that aims to be a trade-off between throughput maximization and satisfying deadlines. There is also a suggestion for M2M uplink transmissions respecting deadlines and the buffer size of M2M users in [6]. In order to satisfy also deadlines of time-critical and heterogeneous CPS data traffic, this paper proposes a traffic-sensitive real-time scheduling algorithm and a performance comparison of the novel algorithm and EDF. In contrast to EDF that considers deadlines exclusively, the solution approach additionally takes the data amount of each user for transmission into account. The combination of these two aspects is introduced as PayDA scheduling scheme for mobile networks with the objective to enhance their efficiency in terms of the compliance with deadlines for a preferably high number of users.

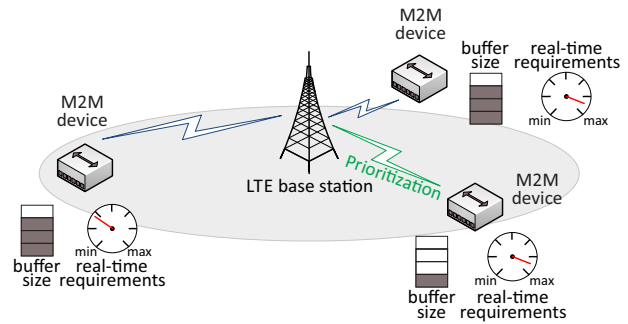

Fig. 1. Example scenario for prioritization of a UE concerning the allocation of spectral radio resources based on temporal requirements regarding the data transmission and the buffer size.

TABLE I
COMPARISON OF DIFFERENT SCHEDULING ALGORITHMS: OBJECTIVES AND SCHEDULING CRITERIA

| Scheduling algorithm | Optimization goals | | | Scheduling criteria | | |
|---|---|---|---|---|---|---|
| | Max. Fairness | Max. Throughput | Adherence to deadlines | Deadlines | Packet size | Channel conditions |
| Round Robin | yes | no | no | no | no | no |
| Maximum Throughput | no | yes | no | no | no | yes |
| Proportional Fair | partial | partial | no | no | no | partial |
| Earliest Deadline First | no | no | yes | yes | no | no |
| Proposed PayDA | partial | partial | yes | yes | yes | no |

### III. PAYLOAD-SIZE AND DEADLINE-AWARE SCHEDULING FOR TIME-CRITICAL CYBER PHYSICAL SYSTEMS

As mentioned in the previous section, the proposed PayDA scheme is an enhancement of the common EDF algorithm and extended by the consideration of the remaining data amount of each data flow. The idea of EDF is formulated as

$$\omega_i = \frac{1}{(\tau_i - D_{HOL,i})} \quad (1)$$

where $\omega_i$ is the weight of a metric used for user prioritization, $\tau_i$ is the deadline and $D_{HOL,i}$ is the head-of-line delay for the $i$-th user. The head-of-line delay $D_{HOL}$ marks the duration of stay of the first packet in a packet queue on the part of the transmitter since its generation. The less time up to a deadline is left the higher is the priority of a packet for transmission. Because EDF was originally designed for scheduling in operating systems, it is less suitable for use in highly dynamic wireless communication systems. Therefore, we extend the EDF's approach to the consideration of the remaining data amount each user wants to receive. Hence, Eq. 1 has to be adapted to

$$\omega_i = \frac{1}{((\tau_i - D_{HOL,i}) \cdot \delta_{left,i})} \quad (2)$$

where $\delta_{left,i}$ is the remaining data amount of the $i$-th user. Obviously, remaining data amount and remaining time are of equivalent significance for the metric calculation. Thus, a data stream with a close deadline and a small data amount tends to be preferred in contrast to data streams with either large data amounts or much time up to their deadlines. The block diagram in Fig. 2 demonstrates the fundamental process steps of PayDA. The main focus of PayDA is on the metric calculation for each RB and data traffic flow within a Transmit Time Interval (TTI) by using the formula of Eq. 2. After the metric calculation is finished, the RBs are allocated iteratively to the corresponding UEs according to their metrics.

In this way, an efficient allocation of RBs concerning the compliance with deadlines to time-sensitive data transmissions can be guaranteed. In order to delimit the functionality of various popular scheduling techniques, Tab. I summarizes the objectives and scheduling criteria of EDF and PayDA as well as of the common best effort scheduling strategies Maximum Throughput (MT), Proportional Fair (PF) and Round Robin (RR).

### IV. SIMULATION-BASED SYSTEM MODEL

In this section, the simulation-based system model that is used for the performance evaluation of PayDA is presented. It consists of the description of the actual LTE simulation environment with a scenario description and further data traffic classes in LTE.

TABLE II
TRANSPORT CLASSES ALONG WITH APPLICATION TYPES EXAMINED IN LTE-SIM

| Data traffic class | Application type |
|---|---|
| H2H | web, video, voice |
| real-time V2X | safety-relevant (e.g., traffic report) |
| | convenience (e.g., multimedia) |

#### A. Data traffic classes in LTE

Further types of mobile applications including various requirements for data transmission have been added to LTE-Sim to reproduce a realistic and heterogeneous mobile data traffic in the simulation scenario. Tab. II provides an overview of the considered data traffic classes and the explicitly analyzed time-critical and time-tolerant applications. While several multimedia and safety-relevant services are among the former application type, applications in the field of CPS (e.g., smart meters) and convenience (e.g., web browsing) mostly belong to the latter group. Every mobile user runs only one application whose start time is randomly chosen from a set of equally distributed values over the entire simulation time.

#### B. LTE Simulation with Scenario Description

In the simulation scenario, we use an urban environment to evaluate the performance of PayDA. The scenario consists of one macro cell with a radius of one kilometre and an increasing number of users using the Manhattan mobility algorithm for movement. Tab. III gives an overview of the most important simulation parameters. In addition, Tab. IV lists the parameters of all involved applications used to generate heterogeneous data traffic. In particular, this data traffic has various time-critical and volume-oriented requirements for its transmission. Finally, the performance of EDF and PayDA should also be examined for a worst case scenario, where the data traffic with almost identical timing requirements but significant miscellaneous payload sizes is generated by two safety-relevant vehicular applications.

TABLE III
PARAMETERS FOR THE SIMULATION OF A REALISTIC URBAN SCENARIO IN LTE-SIM

| Simulation parameter | Value |
|---|---|
| Cell radius | 1 km (Urban) |
| System bandwidth | 20 MHz (100 Resource Blocks) |
| Operating Frequency | 2.6 GHz |
| Channel model | Typical urban 3GPP |
| Max. RF power (Evolved Node B) | 40 dBm |
| Max. RF power (User Equipment) | 23 dBm |
| Number of User Equipments | 50-130 (incrementally increased) |
| Total simulation period | 100 seconds |
| Start times of applications | Equally distributed |
| Mobility | Manhattan mobility |

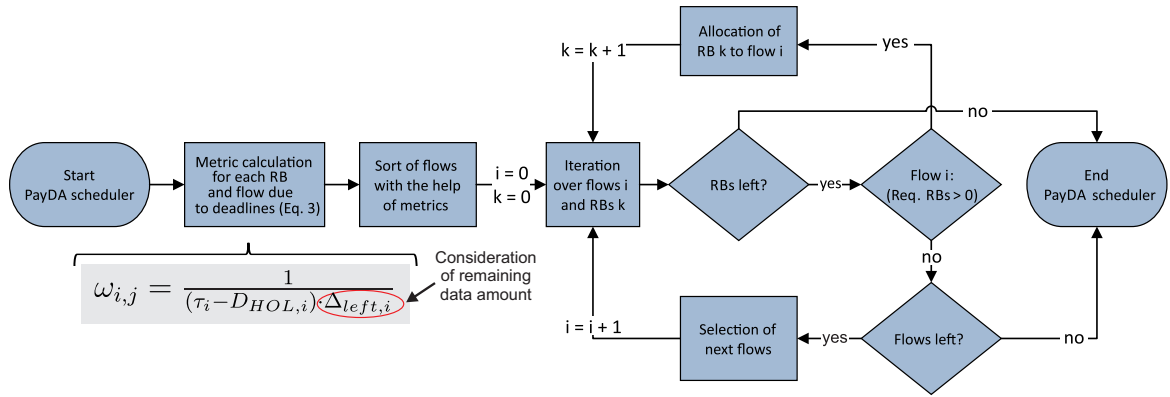

Fig. 2. Block diagram of the iterative process steps of the proposed PayDA packet scheduler: in each iteration the current resource block $k$ is allocated to the user equipment corresponding to the data traffic flow $i$ with the currently highest metric

## V. Results of performance evaluation

We primarily compare the performances of EDF and PayDA: EDF just accounts for the remaining time to a deadline, whereas PayDA also respects the remaining payload size for data transmission. In this regard, two types of data traffic are examined. At first, in Sec. V-A a heterogeneous data traffic is analyzed (Tab. IV), hereinafter the performance of both mechanisms is evaluated for a homogeneous data traffic (cf. Sec. V-B).

### A. Urban scenario with heterogeneous data traffic

Fig. 3 illustrates the performance results of EDF, PayDA and the non-real-time algorithms MT, PF, RR in relation to the DMR and the latency for a heterogeneous data traffic (c.f. Tab. IV). It should be noted that a comparatively high number of 80 mobile users who generate the interfering File Transfer Protocol (FTP) data traffic are taken into account. The left side of Fig. 3 reveals an optimal performance of EDF and PayDA regarding the DMR. In this case, considerably less than 1% of all scheduled data packets cannot be transmitted within the given deadlines. Compared to these values, the maximum DMRs of MT, PF and RR are remarkably high (about 30% up to 60%). The reason for this low performance is a variety of objectives, partially disregarding deadlines completely. On the right hand side of Fig. 3 the latencies of all packet transmissions are illustrated. In contrast to the analogue DMRs, a significant difference of up to 14 seconds regarding the median latencies can be determined for the benefit of PayDA compared to EDF. Even the use of MT leads to an about 9 seconds lower median, whereas the application of RR and PF results in a slightly higher median latency. These distinct latencies of EDF and PayDA result from the heterogeneous patterns of the data traffic. While the most H2H applications generate high-rate and time-critical data traffic with small payload sizes (e.g., Voice over IP (VoIP)), the V2X data traffic is characterized by large and time-tolerant FTP packets. This mixture of patterns could not be handled efficiently by EDF due to its exclusive consideration of deadlines. As long as time-critical data packets are scheduled, all RBs are exclusively used for their transmission. Another advantage of PayDA contrary to EDF is a higher average goodput. Fig. 4 contrasts the differences of these goodputs using EDF, PayDA, MT, PF, RR. Apparently, an increasing number of interfering users has diverging impacts on the average goodputs. In the case of no interfering users (100% time-critical H2H data traffic), the goodputs initially increase when using EDF, PayDA and PF, while they drop continuously beginning at a higher level if MT and RR are used. This stems from the fact that the former procedures primarily allocate RBs

TABLE IV
APPLICATION PARAMETRIZATION USED FOR SIMULATING AN URBAN SCENARIO WITH HETEROGENOUS DATA TRAFFIC IN LTE-SIM [7]

| Application | Max. delay [s] | Payload [Byte] | Interval [s] | Data rate [kbit/s] |
|---|---|---|---|---|
| VoIP | 0.1 | 32 | 0.02 | 12.8 |
| Video (H.264) | 1 | var. | 0.4 | 242 |
| Video call (High Quality) | 0.2 | var. | - | 500 |
| Video call (High Definition) | 0.2 | var. | - | 1500 |
| Music (MP3) | 0.05 | 52 | 0.003 | - |
| Website | 10 | $5 \cdot 10^3$ | - | - |
| File Transfer Protocol | 60 | $5.25 \cdot 10^6$ | 30 | - |

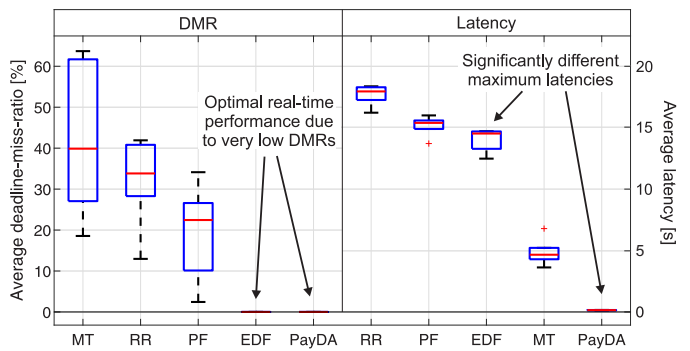

Fig. 3. Heterogeneous data traffic scenario: Comparison of average DMRs and mean latencies for various scheduling strategies. Proposed PayDA achieves lowest average DMR and latency.

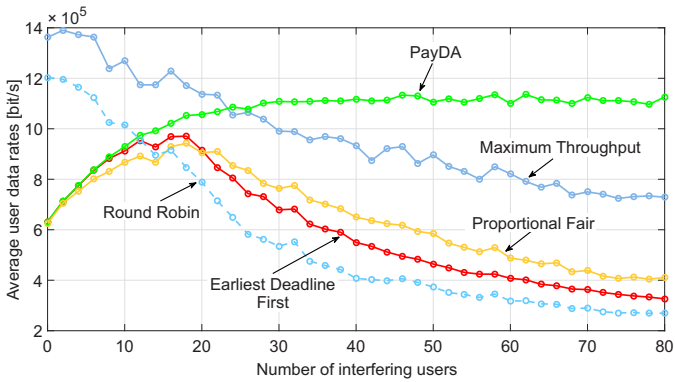

Fig. 4. Heterogeneous data traffic scenario: Analysis of the overall average user goodput for an increasing interfering data traffic when using different scheduling algorithms

to those users generating high-rate, time-critical and small-sized data traffic (e.g., VoIP, MP3 stream), whereas the use of MT and RR also leads to a transmission of time-tolerant and large-sized data packets (e.g., video stream, website). At a number of 80 interfering users, the average goodput when using PayDA is larger by a factor of about 1.5 and 3.5 compared to those of MT and EDF, respectively. This issue is thus, that PayDA also allocates RBs to large-sized FTP packet transmissions, whereas EDF penalizes those data streams because of their less strict deadlines. The more FTP data is scheduled for transmission, the greater the decrease of the mean goodput will be. When using MT and RR, time-critical data transmissions are penalized resulting in decreasing goodputs caused by higher DMRs.

*B. Urban scenario with homogeneous data traffic*

For a nearly homogeneous, time-critical and high-rate data traffic two vehicular applications are used (cf. Tab. V). Fig. 5 shows the DMRs of EDF and PayDA for various start time intervals of all data transmissions respecting different sizes and equally distributed values. Obviously, the DMRs of both schedulers are affected in varying degrees depending on the temporal compactness of the start times. Overall, an increasing dispersion of the start times leads to decreasing DMRs. Nevertheless, large differences in the performance are revealed particularly for small intervals. At this point, PayDA is considerably predominant with a gain by more than 35% with regard to the maximum DMR. With an increasing size of the start times interval the difference between the maximum DMRs is reduced to 25% (1 second interval) or is even negligible (100 seconds interval), respectively. The reason for this decreasing divergence of the maximum DMRs results from a lower density of data traffic, because some applications (e.g., web browsing, FTP) only need a specific time period for data transmission.

TABLE V
APPLICATION PARAMETERS FOR AN URBAN SCENARIO WITH TEMPORALLY HOMOGENEOUS DATA TRAFFIC IN LTE-SIM

| Application | Max. delay [s] | Payload [Byte] | Interval [s] | Data rate [kbit/s] |
|---|---|---|---|---|
| Context Awareness Messages | 0.1 | 100 | 0.1 | 1 |
| Future time-critical application | 0.1 | 10000 | 1 | 10 |

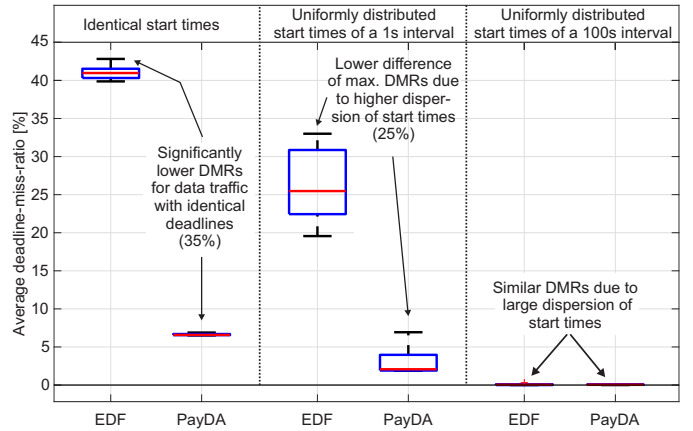

Fig. 5. Homogeneous data traffic scenario: Performance of PayDA in relation to EDF when using different start time intervals

## VI. CONCLUSION

In this paper, we have quantified the performance of various scheduling algorithms regarding the compliance with deadlines for different types of data traffic with the help of a complex LTE simulation environment. In addition, we proposed the efficiency and performance improved real-time scheduling scheme PayDA. PayDA seizes the concept of EDF but extends it by the consideration of remaining payload sizes. The increase of efficiency and performance has been evaluated by a simulation-based system model. According to the results, PayDA leads to significantly lower latencies and higher goodputs at comparable low DMRs for heterogeneous data traffic. In the case of time homogeneous data traffic, especially the start times of applications are relevant. At this juncture, PayDA clearly predominates for very similar start times of various data transmissions. In the future, we will evaluate PayDA for further types of data traffic.

ACKNOWLEDGMENT

Part of the work on this paper has been supported by Deutsche Forschungsgemeinschaft (DFG) within the Collaborative Research Center SFB 876 "Providing Information by Resource-Constrained Analysis", project B4.

REFERENCES

[1] G. Piro, L. A. Grieco, G. Boggia, F. Capozzi, and P. Camarda, "Simulating LTE cellular systems: an open-source framework," *IEEE Transactions on Vehicular Technology*, vol. 60, no. 2, pp. 498–513, Feb 2011.
[2] F. Capozzi, G. Piro, L. A. Grieco, G. Boggia, and P. Camarda, "Downlink packet scheduling in LTE cellular networks: key design issues and a survey," *IEEE Communications Surveys Tutorials*, vol. 15, no. 2, pp. 678–700, Second 2013.
[3] G. Piro, L. A. Grieco, G. Boggia, R. Fortuna, and P. Camarda, "Two-level downlink scheduling for real-time multimedia services in LTE networks," *IEEE Transactions on Multimedia*, vol. 13, no. 5, pp. 1052–1065, Oct 2011.
[4] C. Ide, B. Dusza, and C. Wietfeld, "Client-based control of the interdependence between LTE MTC and human data traffic in vehicular environments," *IEEE Transactions on Vehicular Technology*, vol. 64, no. 5, pp. 1856–1871, May 2015.
[5] A. Elhamy and Y. Gadallah, "BAT: A balanced alternating technique for M2M uplink scheduling over LTE," in *2015 IEEE 81st Vehicular Technology Conference (VTC Spring)*, May 2015, pp. 1–6.
[6] N. Afrin, J. Brown, and J. Y. Khan, "Performance analysis of an enhanced delay sensitive LTE uplink scheduler for M2M traffic," in *Telecommunication Networks and Applications Conference (ATNAC), 2013 Australasian*, Nov 2013, pp. 154–159.
[7] M. Kihl, K. Bür, P. Mahanta, and E. Coelingh, "3GPP LTE downlink scheduling strategies in vehicle-to-infrastructure communications for traffic safety applications," in *Computers and Communications (ISCC), 2012 IEEE Symposium on*, July 2012, pp. 000 448–000 453.